\documentclass[%
preprint, 
superscriptaddress,
amsmath,amssymb,
aps, physrev,
]{revtex4-2}

\usepackage[english]{babel}
\usepackage{graphicx}
\usepackage{dcolumn}
\usepackage{bm}
\usepackage{caption}
\usepackage{subcaption}

\begin{document}

\preprint{APS/123-QED}

\title{\textbf{Brittle-to-ductile transition and strain relaxation in Si$_{1-x}$Ge$_x$ linearly graded buffers}} 

\author{Riccardo Civiero}
\affiliation{LNESS Dipartimento di Fisica, Politecnico di Milano, via F. Anzani 42, 22100 Como, Italy.}

\author{Elena Campagna}
\affiliation{Dipartimento di Scienze, Università degli Studi Roma Tre, Viale G. Marconi 446, Roma 00146, Italy}

\author{Afonso Cerdeira Oliveira}
\affiliation{LNESS Dipartimento di Fisica, Politecnico di Milano, via F. Anzani 42, 22100 Como, Italy.}

\author{Marvin Hartwig Zoellner}
\affiliation{IHP - Leibniz Institute for High Performance Microelectronics, Frankfurt (Oder) 15236, Germany}

\author{Davide Impelluso}
\affiliation{LNESS Dipartimento di Fisica, Politecnico di Milano, via F. Anzani 42, 22100 Como, Italy.}

\author{Daniel Chrastina}
\affiliation{LNESS Dipartimento di Fisica, Politecnico di Milano, via F. Anzani 42, 22100 Como, Italy.}

\author{Giovanni Capellini}
\affiliation{Dipartimento di Scienze, Università degli Studi Roma Tre, Viale G. Marconi 446, Roma 00146, Italy}
\affiliation{IHP - Leibniz Institute for High Performance Microelectronics, Frankfurt (Oder) 15236, Germany}

\author{Giovanni Isella}

\affiliation{LNESS Dipartimento di Fisica, Politecnico di Milano, via F. Anzani 42, 22100 Como, Italy.}

\date{\today}

\begin{abstract}
 The strain-relaxation mechanism of a set of Si$_{0.6}$Ge$_{0.4}$ linearly graded buffers (LGBs), grown following different temperature profiles, has been investigated by means of defect-etching and variable-temperature high-resolution X-ray diffraction (VT-HRXRD). Defect-etching experiments demonstrate that a sharp increase of threading dislocation density (TDD) from $3 \times 10^{5}$\,cm$^{-2}$ to $1.2 \times 10^{6}$\,cm$^{-2}$ takes place when the final growth temperature exceeds a critical value T$_c\approx 530^\circ$C. VT-HRXRD measurements show that in low TDD samples extra relaxation takes place for annealing temperatures larger than  T$_c$, thanks to the nucleation of new dislocations. These results indicate that, below T$_c$, strain relaxation is driven by the gliding of existing dislocations while above T$_c$ new dislocations are nucleated, suggesting a link with our results and the brittle-to-ductile transition in Si$_{1-x}$Ge$_x$ alloys.  
\end{abstract}


\maketitle 

\section{Introduction}
Strain-relaxed buffer-layers (SRBs) play a key-role in hetero-epitaxy, enabling the deposition of strained-engineered active layers on commercially available substrates.  An ideal SRB should plastically accommodate the lattice mismatch between substrate and active layer, through the introduction of misfit dislocations (MDs), yet minimizing the density of threading dislocations (TDs) which degrades the electro-optical properties of the device layer \cite{giovane_correlation_2001}. 

In constant composition SRBs, misfit segments accumulate at the buffer/substrate interface, hindering the glide of existing dislocations and promoting the nucleation of new dislocations. As a result, in the Si$_{1-x}$Ge$_x$/Si system, constant composition layers features a TD density (TDD) exceeding $10^7$\,cm$^{-2}$ already for $x \approx 0.3$ and monotonically increasing with Ge content \cite{bolkhovityanov_artificial_2003}.

More than 40 years ago Abrahams \emph{et. al.} introduced linearly graded buffers (LGBs) in GaAs$_{1-x}$P$_x$/GaAs hetero-epitaxy with the aim of reducing the TDD, as compared to constant composition buffers \cite{abrahams_dislocation_1969}.
Later, LGBs  were extensively used in the early days of strained-Si/Si$_{1-x}$Ge$_x$ technology \cite{schaffler_strained_1994} before being outperformed by local-stressor approaches \cite{thompson_90-nm_2004}, however, they recently attracted renewed interest in the framework of semiconductor-based spin qubits \cite{neyens_probing_2024, SHIMURA2024108231, jirovec_singlet-triplet_2021} and integrated photonics in the near and mid infrared \cite{marris-morini_germanium-based_2018,Vakarin20173482, Frigerio20213573}.  
 
In Si$_{1-x}$Ge$_x$ LGBs the alloy composition $x$ is linearly increased from $x=0$ to the final target composition $x_f$ and then terminated by a constant composition Si$_{1-x_f}$Ge$_{x_f}$ layer.  This approach offers several advantages when compared to constant composition SRBs. Misfit dislocations are not accumulated at the interface, but distributed throughout the linearly graded region, strongly reducing dislocation pinning. In addition, the formation of misfit segments can take place through the gliding of existing TDs avoiding the nucleation of new dislocations. This led  Abrahams \emph{et. al.} to predict the existence of a minimal TDD capable of relaxing LGBs independently of the final composition. Yet, extensive work carried out on Si$_{1-x}$Ge$_x$ by Fitzgerald and coworkers  \cite{fitzgerald_dislocation_1999, samavedam_novel_1997} showed that LGBs grown by chemical vapor deposition (CVD) features a TDD increase with increasing Ge content, highlighting the role played by processes other than glide, such as nucleation and multiplication, in the relaxation process \cite{isaacson_deviations_2006}.

Two different trends are expected for the temperature dependence of TDD in glide-driven or nucleation-driven relaxation \cite{isaacson_deviations_2006}. As T increases, dislocation glide becomes more effective requiring a comparatively low TDD to promote strain relaxation. In contrast, in nucleation-driven relaxation, a higher T favors the nucleation of more dislocations and an increased TDD. 

Most of the experimental works reported so far on Si$_{1-x}$Ge$_x$ LGBs with $x_f\leq 0.5$ are focused on epilayers deposited by CVD in the 650-1000$^\circ$C temperature range \cite{fitzgerald_dislocation_1999,Bogumilowicz2006,zoellner}. With the aim of studying the competition between glide and nucleation driven relaxation in a so far unexplored temperature range, we have investigated by means of T-dependent high-resolution X-Ray diffraction (T-HRXRD), atomic force microscopy (AFM), and etch-pit counting (EPC), a set of Si$_{1-x}$Ge$_x$ LGBs with $x_f=0.4$ grown at different final temperatures comprised between 510 and 580$^\circ$C. Interestingly, the switch from glide-driven to nucleation-driven relaxation takes place in a narrow  temperature range around T$_c \approx 530^\circ$C. The analysis of additional LGBs with $x=0.1, 0.2, 0.3$  confirms the dominance of glide-driven relaxation for samples grown at lower temperatures.       
       
\section{Epitaxial growth and characterization}
All samples analyzed in this work were deposited on 100 mm Si(001) wafers by low-energy plasma-enhanced CVD \cite{rosenblad:1998:5} (LEPECVD) using SiH$_4$ and GeH$_4$ precursors. In LEPECVD the activation of process gases is performed by means of a discharge plasma \cite{rondanini_experimental_2008}, whose density can be tuned to vary the deposition rate between 0.5\,\r{A}/s and 10\,nm/s, independently on the substrate temperature. In this work an average deposition rate of $\approx$\, 8 nm/s was used. The linear compositional profile was approximated by increasing the Ge concentration in the epilayer in $0.5\%$ steps, \cite{sanchez-almazan_ge_2004} to obtain a $7\%/\mu$m grading rate. The growth was terminated by a 2\,$\mu$m thick constant composition layer. The samples analyzed in this work are listed in Tab.~\ref{tab:samples}.
\begin{table}
	\caption[List of the analyzed samples]{List of the samples grown for this work, reporting: the sample label, the nominal final Ge content and that experimentally measured by X-ray diffraction, the final growth temperature, the degree of relaxation and surface roughness.}
	\label{tab:samples}
	\centering
	\begin{tabular}{c c c c c c}
		\hline
		Sample &  Nom. $x_f$ & Meas. $x_f$ & $T_f$       & Relax. & Rough. \\
  			   &  (\%)       & (\%)   	   & ($^\circ$C) & (\%) & (nm)     \\
		\hline
		A1 & 40 & 39.8 & 510 & 96.9 & 3.12\\
		A2 & 40 & 40.1 & 520 & 97.3 & 2.76\\
		A3 & 40 & 39.9 & 530 & 98.6 & 2.65\\
		A4 & 40 & 39.9 & 550 & 99.3 & 2.52\\
		A5 & 40 & 39.8 & 580 & 100.2 &  2.25\\
		\hline 
		B1 & 30 & 29.5 & 580 & 97.9 & 1.78\\ 
		B5 & 30 & 29.8 & 627 & 100.1 & 1.21\\
		\hline
		C1 & 20 & 19.6 & 650 & 98.7 & 1.24\\  
		C5 & 20 & 19.8 & 673 & 100.5 & 1.11\\
		\hline
		D  & 10 & 9.8  & 720 & 102.2 & 1.67 \\
		\hline
	\end{tabular}
\end{table}

\begin{figure}[t]
	\centering
\includegraphics[width=0.45\textwidth]{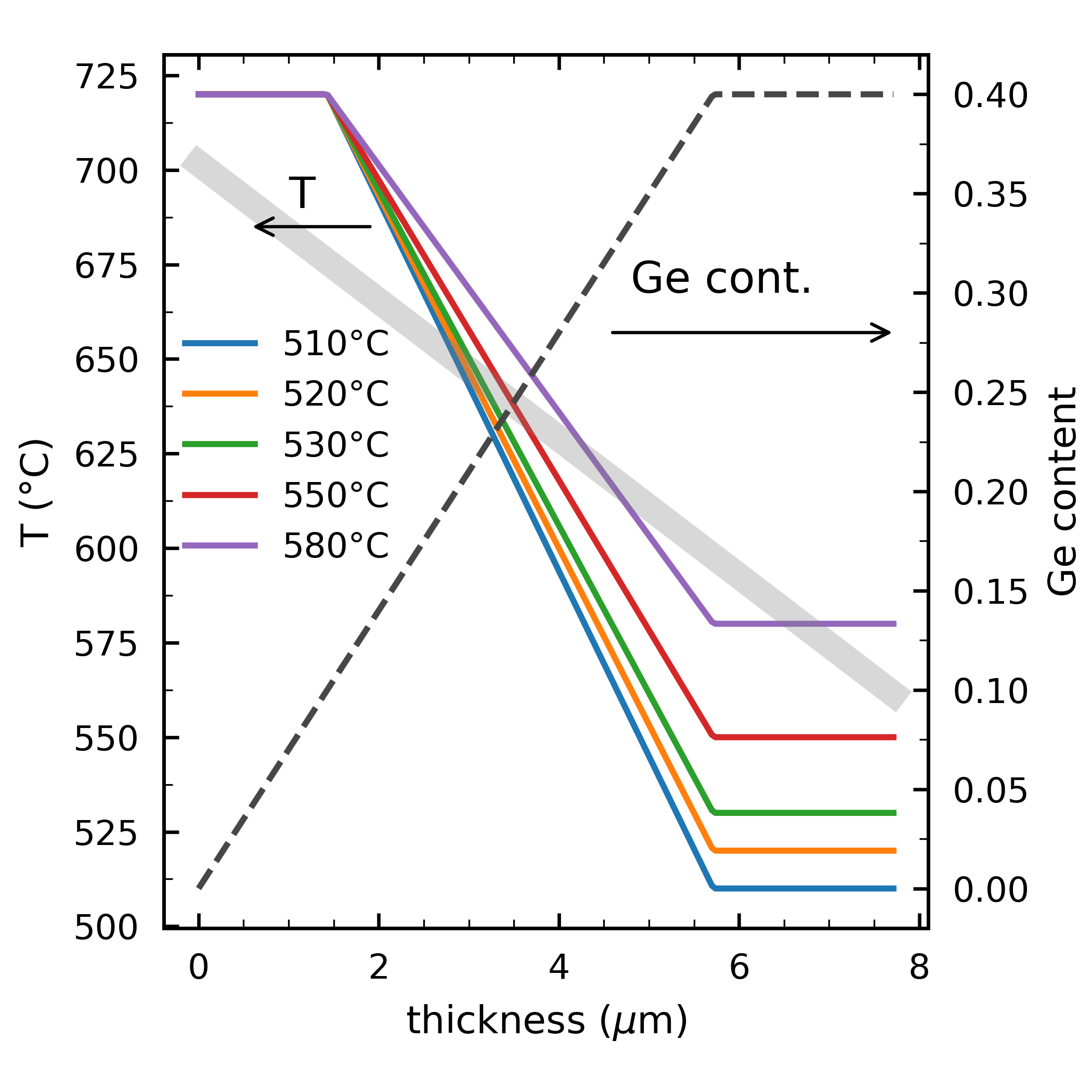}%
	\caption{\label{fig:TProfile} Variation of temperature (left axis) and composition (right axis) during growth for the set of LGBs featuring ${x_f=0.4}$ and different final temperatures. The gray area shows an estimation the brittle -ductile transition temperature in Si$_{1-x}$Ge$_x$.}
\end{figure}

A first set of  Si$_{1-x}$Ge$_{x}$ LGBs with a final Ge content $x_f=0.4$ , labeled A1 to A5, was grown keeping the substrate temperature at T=720$^\circ$C until a Ge content $x=0.1$ was reached. Subsequently,  T was linearly decreased reaching a different final temperature T$_f$ for each one  of the samples under consideration, as shown in Fig.\,\ref{fig:TProfile}.  At a first glance, a set of isothermally grown epilayers might seem a better choice for the study of T-dependent relaxation mechanism. Yet, in Si and Ge diffusion coefficients, vacancy concentrations  \cite{vanhellemont_brother_2007} and dislocation velocity \cite{chaudhuri_velocities_1962} linearly scale with the respective melting points, making the physical mechanism relevant at a given temperature quite different in Si-rich or Ge-rich alloys. A temperature profile decreasing with the Ge content $x$ should, instead, provide a similar energy landscape throughout the LGB growth. In addition, the employed variable-temperature growth protocol kinetically suppress surface roughening, which might impact dislocation nucleation and glide, \cite{samavedam_novel_1997} without the need of chemical-mechanical polishing steps commonly employed in thermal-CVD grown LGBs.

LGBs with $x_f=0.3$ (B1 and B5), $x_f=0.2$ (C1 and C5) and $x_f=0.1$ (D) were also deposited. The T profile of samples B1 (B5) overlaps that of sample A1 (A5) in the common compositional range (\emph{i. e. } $x \leq 0.3$). In the same way samples C1 and C5 share the same T profile of A1 and A5 for $x \leq 0.2$. In short, samples B and C can be considered as ``snapshots'' of the corresponding A-samples at intermediate points ($x=0.3,0.2$) of the growth. The series is completed by a LGB with $x_f=0.1$ deposited at a constant T=720$^\circ$C.

A Rigaku SmartLab diffractometer, equipped with a DHS1100 oven from Anton Paar was used for temperature variable  high-resolution X-ray diffraction (VT-HRXRD) measurements \cite{capellini_high_2012} performed in N$_2$ atmosphere to avoid surface oxidation. The residual in-plane strain $\epsilon_\parallel$ and final Ge content $x_f$ of the LGBs were determined using reciprocal space maps (RSMs) of the (004) and steep-incidence (224) reflections. The degree of relaxation $\beta=1-\epsilon_{\parallel}/f$  was obtained by calculating the lattice mismatch $f$, between the Si$_{1-x_f}$Ge$_{x_f}$ top layer and the Si substrate, using the quadratic interpolation of the Si$_{1-x}$Ge$_{x}$ lattice parameter reported in Ref. \cite{Dismukes1964}.  

\captionsetup[subfigure]{
	position=bottom,
	skip=-20pt,
	singlelinecheck=false,
	justification=raggedright
}
\begin{figure}[b]
	\centering
	\begin{subfigure}[b]{0.45\textwidth}
		\centering
		\includegraphics[width=\textwidth]{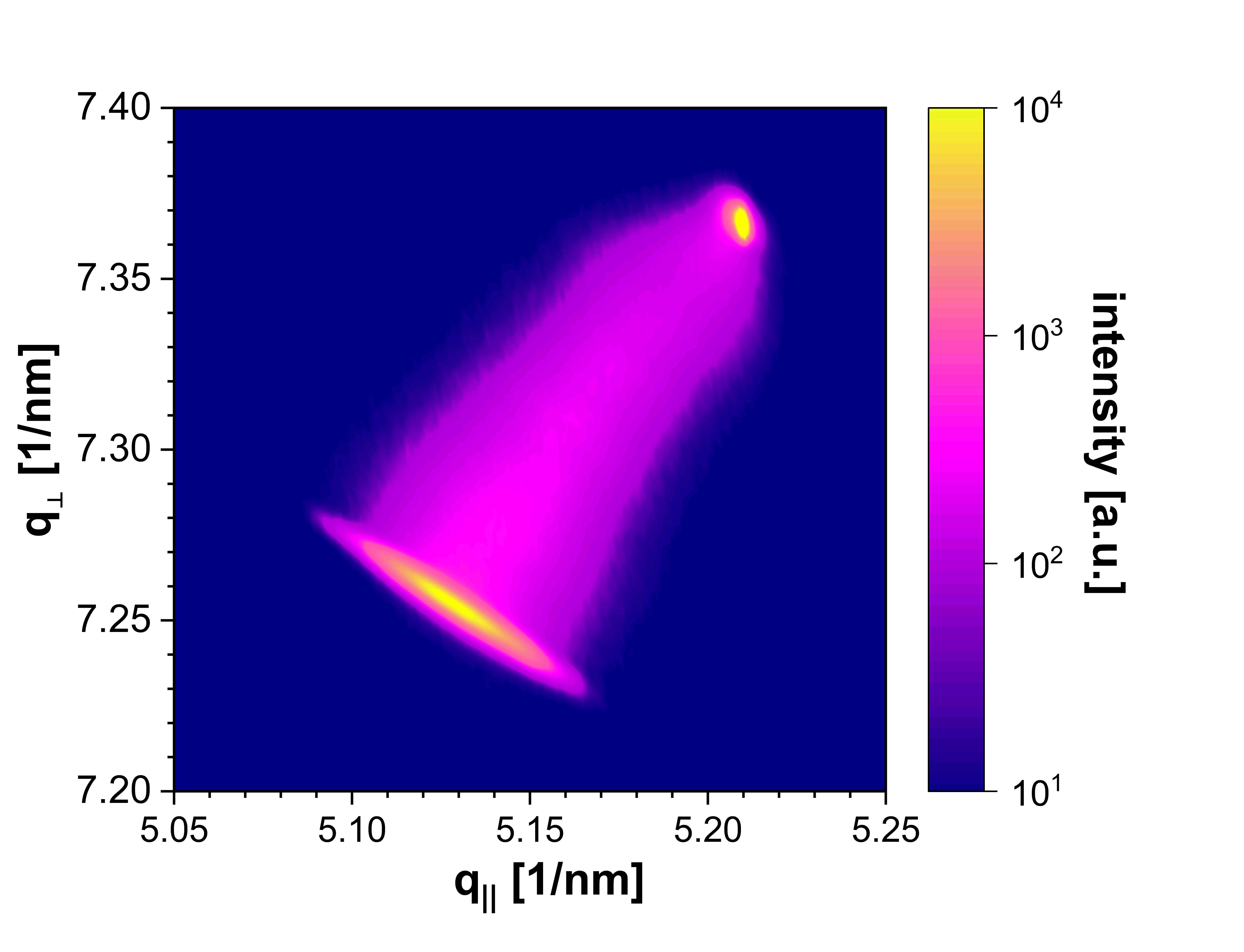}
		\caption{}
	\end{subfigure}
	\vspace{24pt}
	\begin{subfigure}[b]{0.45\textwidth}
		\centering
	\includegraphics[width=0.75\textwidth]{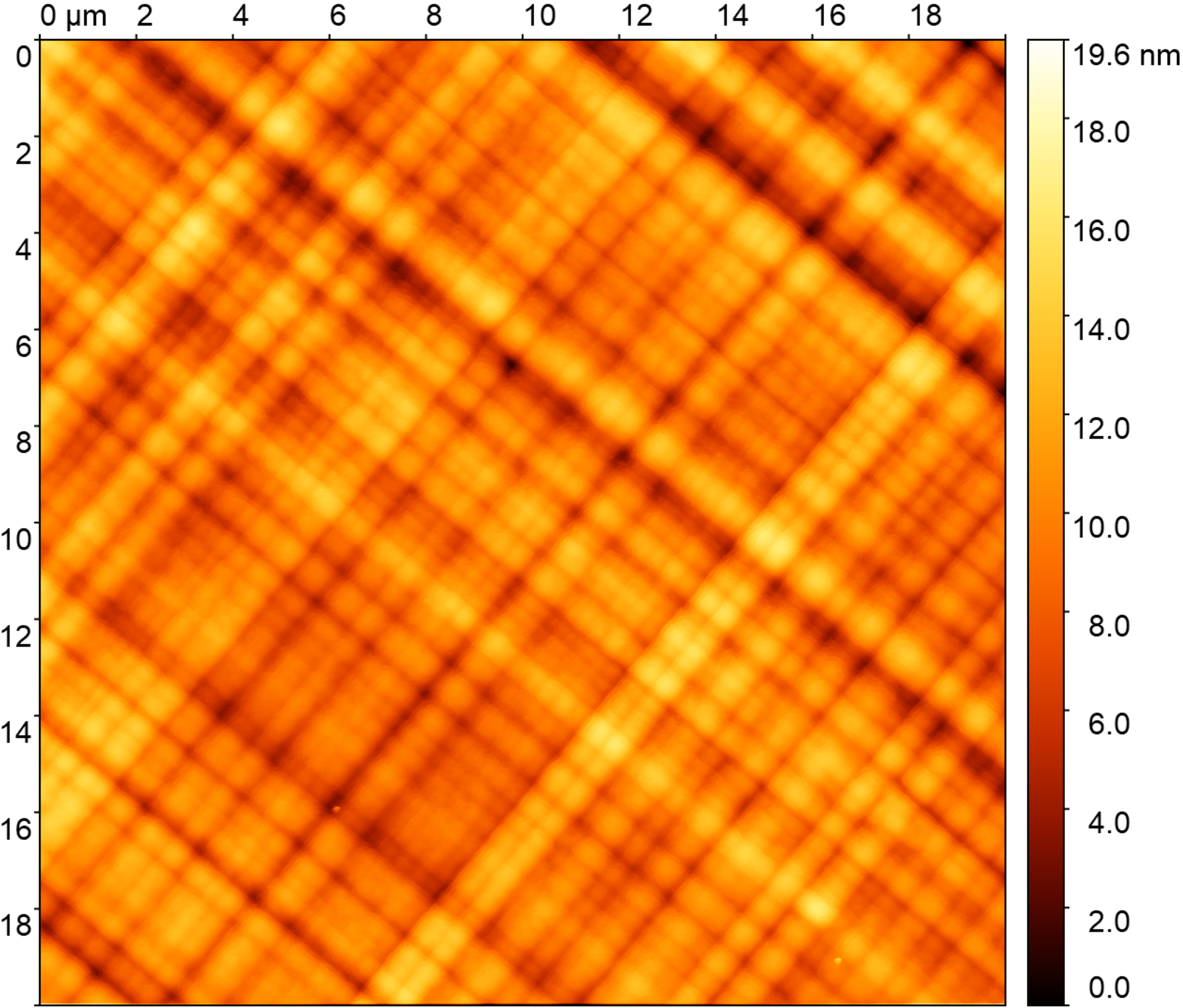}
		\caption{}
	\end{subfigure}
	\caption{\label{fig:xrd} Structural characterization of sample A5. a) Room temperature RSM around the (224) diffraction peak showing the continuous variation of composition between the Si substrate and the Si$_{0.6}$Ge$_{0.4}$ relaxed buffer. b) AFM image of the cross-hatch pattern.}
\end{figure}
Defect etching was carried out using two different etching solutions: the Secco etch and the cold Schimmel etch \cite{Marchionna2006}. Etch-pit counting was performed using  differential interference contrast (DIC) optical microscopy averaging over 10 field-of-view each one covering a $260\,\mu$m $\times 120\,\mu$m region. Atomic force microscopy (AFM) was employed to determine the root mean square (RMS) roughness of the sample surface, averaging over ten  20\,$\mu$m\,$\times$20\,\,$\mu$m scans.

\section{Results and discussion}
The Ge content $x_f$ and degree of relaxation, obtained from RSMs similar to that shown in Fig.\,\ref{fig:xrd}a), are reported in Tab. \ref{tab:samples}. It can be noticed that the measured Ge content matches, within the experimental error, the targeted $x_f$, independently on the T-profile employed for the deposition. The degree of relaxation is, instead, affected by the growth temperature, with samples grown at higher temperatures being more relaxed. Over relaxation is observed for samples A5, B5, C5 and D, indicating that tensile-strain, originating from the thermal mismatch between Ge and Si, contributes to the measured room-temperature residual strain  \cite{capellini_high_2012}.

\begin{figure}[b]
	\centering
	\begin{subfigure}[b]{0.45\textwidth}
		\centering
		\includegraphics[width=\textwidth]{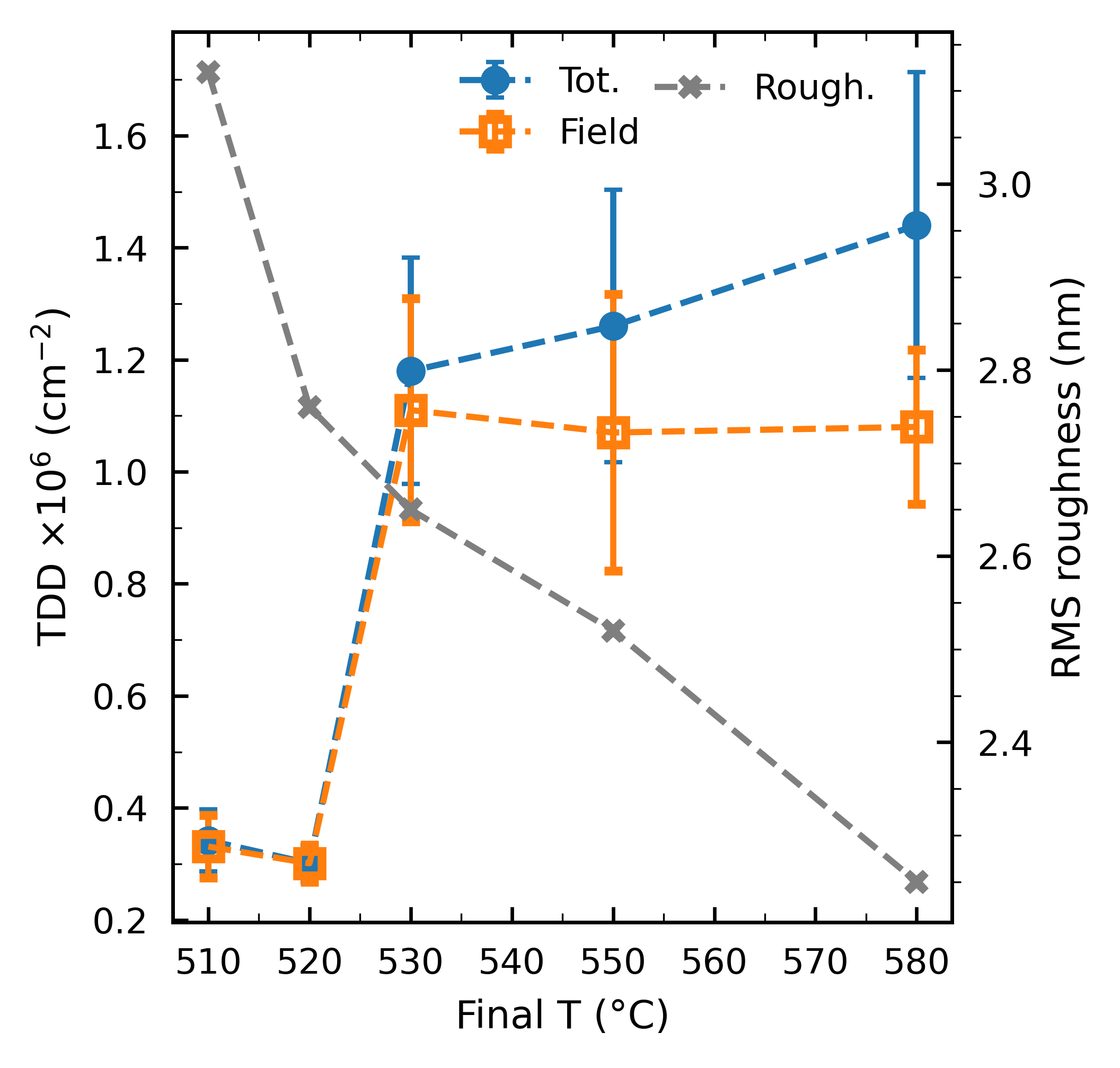}
		\caption{}
	\end{subfigure}
	\begin{subfigure}[b]{0.45\textwidth}
		\centering
		\begin{subfigure}[b]{0.45\textwidth}
			\centering
			\includegraphics[width=\textwidth]{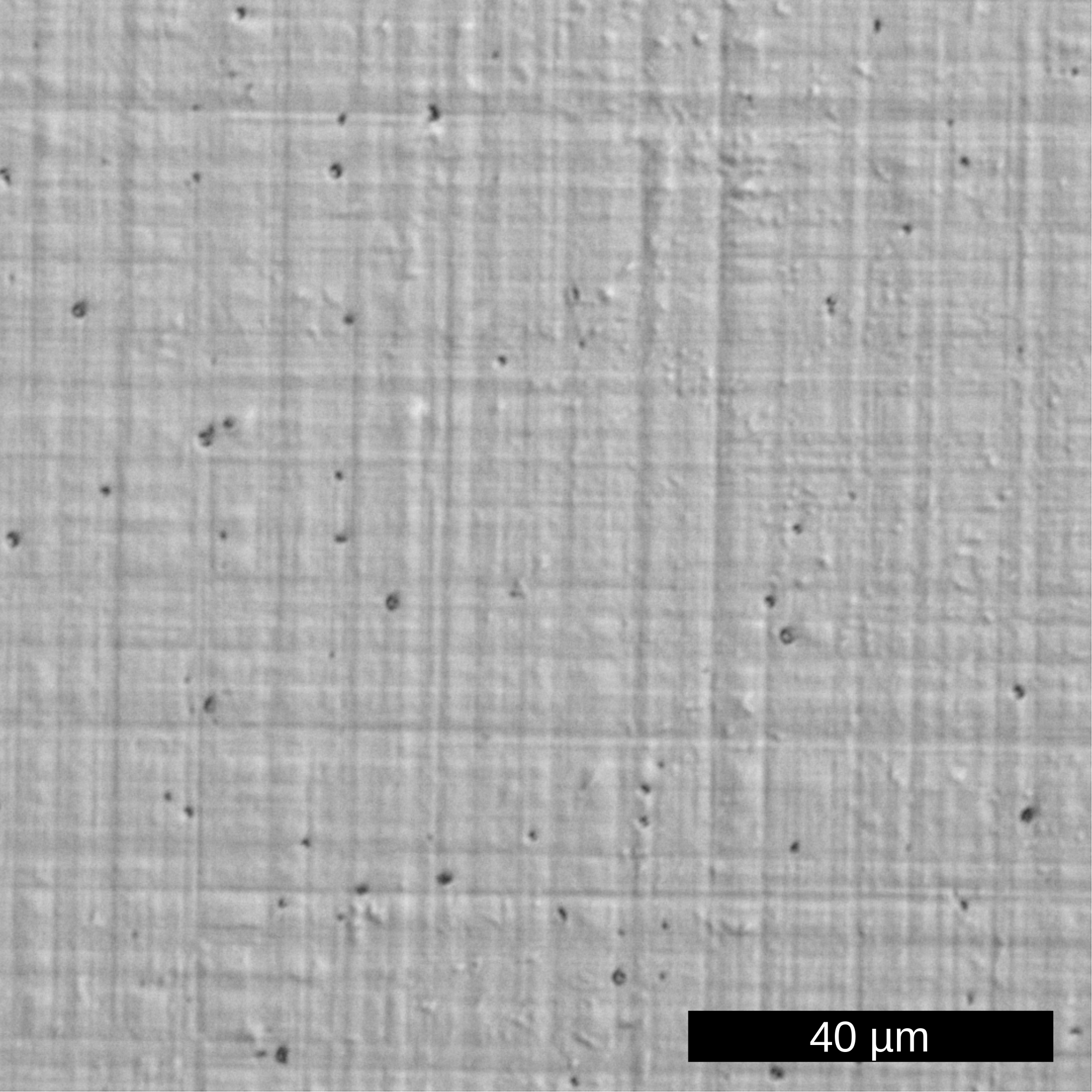}
			\caption{}
		\end{subfigure}
        \begin{subfigure}[b]{0.45\textwidth}
			\centering
			\includegraphics[width=\textwidth]{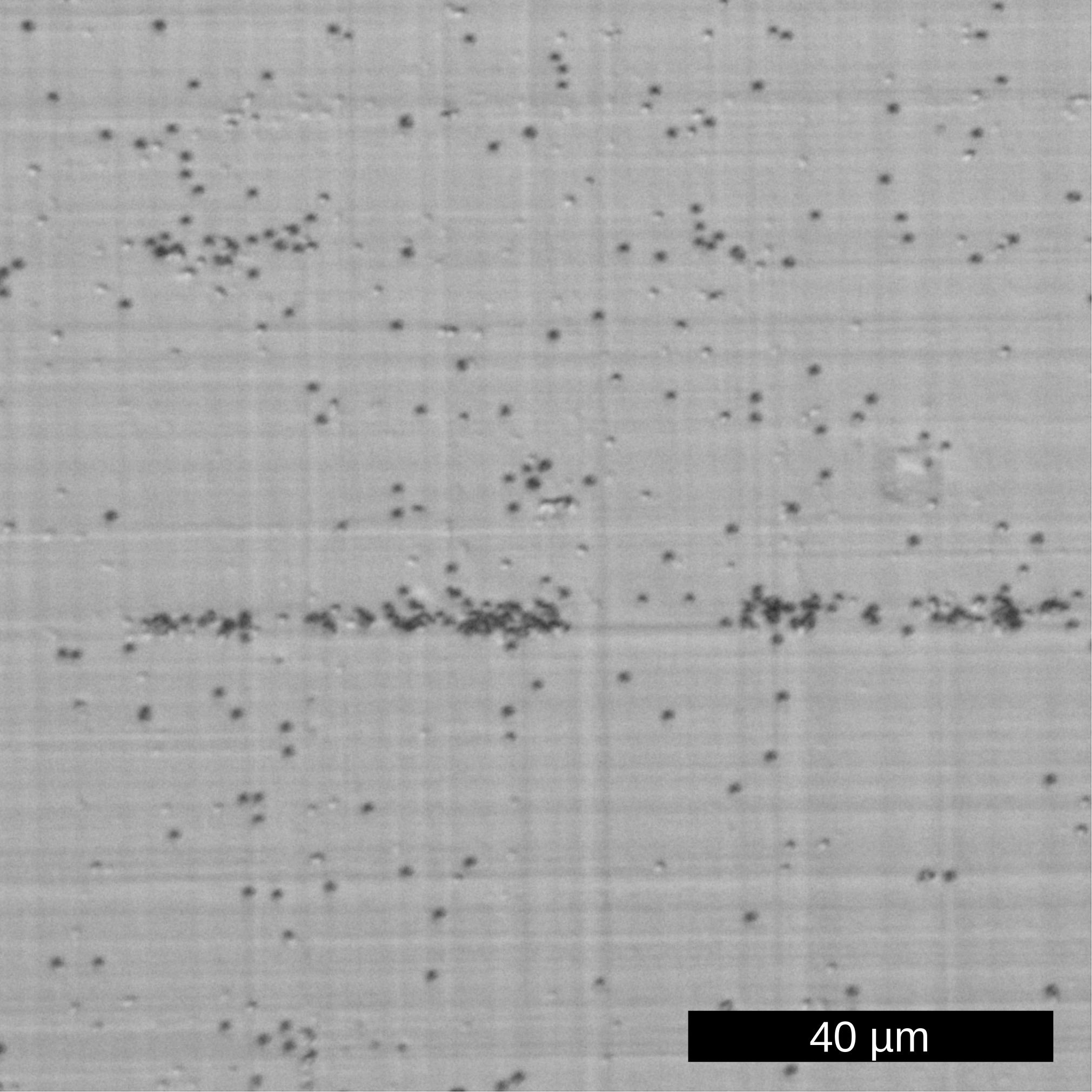}
			\caption{}
		\end{subfigure}
	\end{subfigure}
	\caption{\label{fig:TDD40} Threading dislocation density in Si$_{0.6}$Ge$_{0.4}$ LGBs. (a) TDD \emph{vs.} final temperature. The plot reports the density of so-called field TDs, \emph{i.e.} dislocations not piled-up in clusters, and the total number of TDs. The roughness dependence on the final temperature can be read on the right-hand \emph{y}-axis. Panels (b) and (c) show representative optical micro-graphs obtained after defect etching of samples A1 (T$_f=510\,^\circ$C) and A5 (T$_f=580\,^\circ$C), respectively.}
\end{figure}

All Si$_{0.6}$Ge$_{0.4}$ LGBs are characterized by a cross-hatch pattern similar to that shown in Fig.\,\ref{fig:xrd}b, with an RMS-roughness decreasing with temperature from $3.12$\,nm to $2.25$\,nm. In previous literature reports \cite{samavedam_novel_1997}, a direct correlation between roughness and TDD was observed in thermal-CVD samples where, however, the RMS roughness typically exceeds 100\,nm. In our case, the RMS roughness is relatively low in all samples and, as shown in Fig.\, \ref{fig:TDD40}, features a trend opposite to that of measured TDD, ruling out a possible impact of surface roughness on plastic-relaxation.
\captionsetup[subfigure]{
	position=bottom,
	skip=2pt,
	singlelinecheck=false,
	justification=raggedright
}

Etch-pit counting results for the  Si$_{0.6}$Ge$_{0.4}$ series (samples A1-5 in Tab.\,\ref{tab:samples}) are shown in Fig.\,\ref{fig:TDD40}\,a. It can be noticed that: (a) the TDD sharply increases from $\sim 3 \times 10^{5}$\,cm$^{-2}$ to  $\sim 1.2 \times 10^{6}$\,cm$^{-2}$ varying T$_f$ from 520 to 530$^\circ C$; (b) Low-T samples (A1 and A2) do not feature any statistically relevant density of dislocation pile-ups which, instead, contribute substantially to the total TDD in high-T samples (A3 to A5).

\begin{figure}[t]
	\centering
	\includegraphics[width=0.45\textwidth]{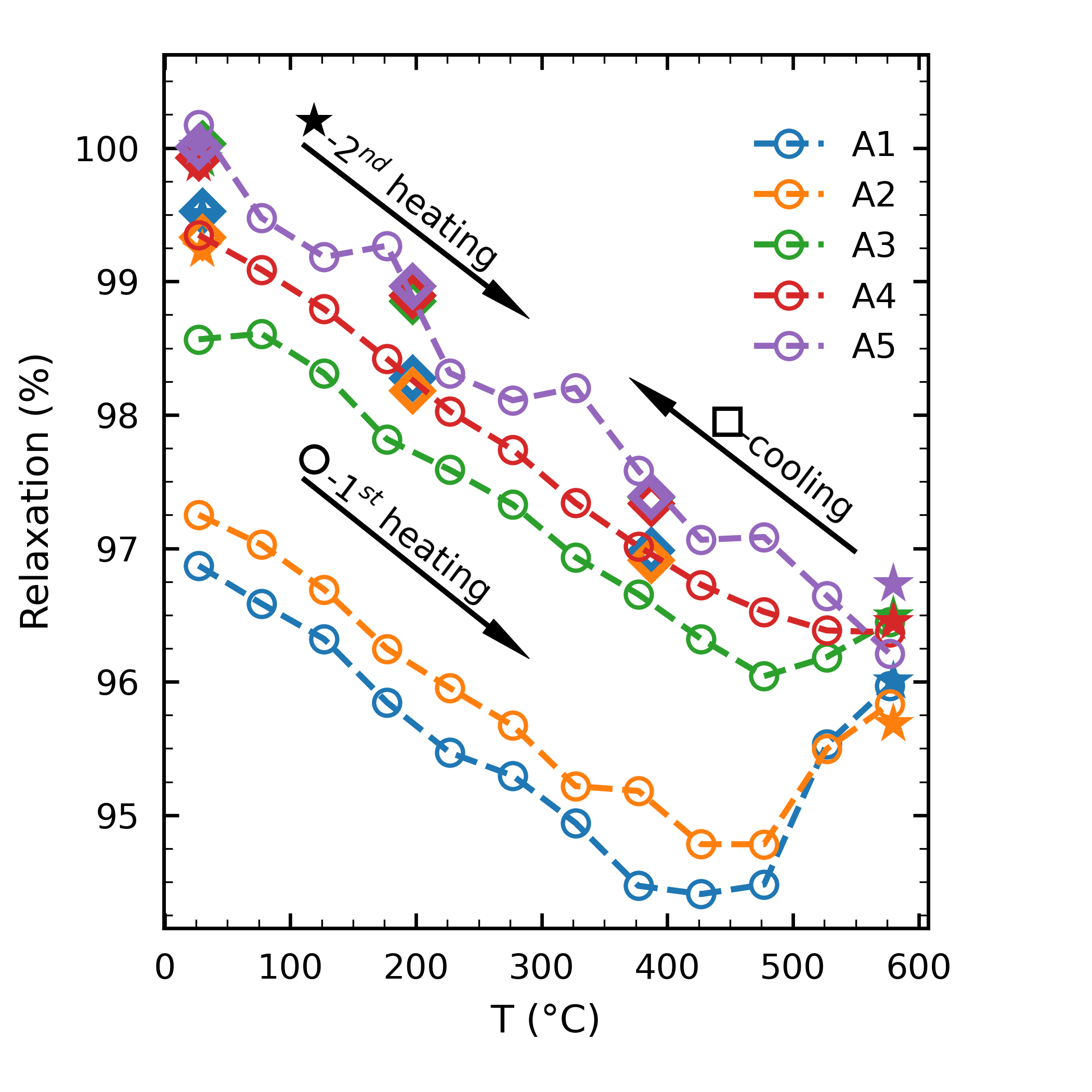}%
	\caption{\label{fig:relaxation} Degree of relaxation at different post-growth annealing temperatures as measured by VT-HRXRD. For each of the A-samples ($x_f=0.4$) listed in Tab.\,\ref{tab:samples} the relaxation has been measured during a 1st heating ramp (empty circles), cool-down (empty diamonds) and a 2nd heating ramp and cool-down (stars)} 
\end{figure}

The as-grown Si$_{0.6}$Ge$_{0.4}$ LGBs were analyzed by means of VT-HRXRD from T=27$^\circ$C to T=557$^\circ$C in 50$^\circ$C steps. The degree of relaxation, measured at every step is indicated by the empty circles in Fig.\,\ref{fig:relaxation}. Subsequently the samples were cooled down re-measuring the degree of relaxation at T= 387, 197 and 27$^\circ$C (empty-diamonds in Fig.\,\ref{fig:relaxation}). Eventually the samples were heated again at T=557$^\circ$C and cooled down at T=27$^\circ$C (filled stars in Fig.\,\ref{fig:relaxation}).  
It can be noticed that, during the first heating cycle, the degree of relaxation decreases for all samples up to T$\approx$500$^\circ$C. As T increases, the (tensile) thermal-strain, accumulated during cool-down after the deposition, is progressively reduced, unveiling the residual (compressive) mismatch-strain, which, not surprisingly, is larger for samples deposited at an average lower-T.

For T $\succsim$ 500$^\circ$C two distinct behaviors are observed for samples A1-A3 and A4-A5, respectively.
In samples A4 and A5 the relaxation measured during the different heating and cooling cycles reversibly follows the same path, indicating that plastic relaxation has reached the equilibrium value achievable for T=577$^\circ$C in the case of Si$_{0.6}$Ge$_{0.4}$ LGBs deposited following the T$_f=580^\circ$C and T$_f=550^\circ$C temperature profile. For these samples the T-dependent degree of relaxation can be attributed solely the to the elastic, thermal-induced strain.

The relaxation of samples A1-A3 is, instead,characterized by a clear step taking place between T$\approx500^\circ$C and T$\approx550^\circ$C, showing that an additional relaxation of misfit-strain is obtained when annealing these samples above their final deposition temperature T$_f=510^\circ$C, T$_f=520^\circ$C and T$_f=530^\circ$C, respectively. Such extra-relaxation is clearly plastic, as indicated by the non-reversible relaxation path followed during cool-down. The second annealing step confirms that a stable configuration, although at slightly different relaxation levels, has been reached for all samples after the first annealing cycle.

In summary the increase of TDs for T$\sim$520-530$^\circ C$ (see Fig.\,\ref{fig:TDD40}) is mirrored by a sudden increase of plastic relaxation observed by VT-HRXRD in a similar temperature range as detailed in Fig.\,\ref{fig:relaxation}. Etch-pit counting performed on the annealed samples yields TDD in the $1-3\times 10^6$\,cm$^{-2}$ range for all samples, demonstrating that, during post-growth annealing, the extra relaxation takes place mainly by the nucleation of new dislocation and not by gliding of existing TDs.
    \begin{figure}[t]
    	\centering
    	\includegraphics[width=0.45\textwidth]{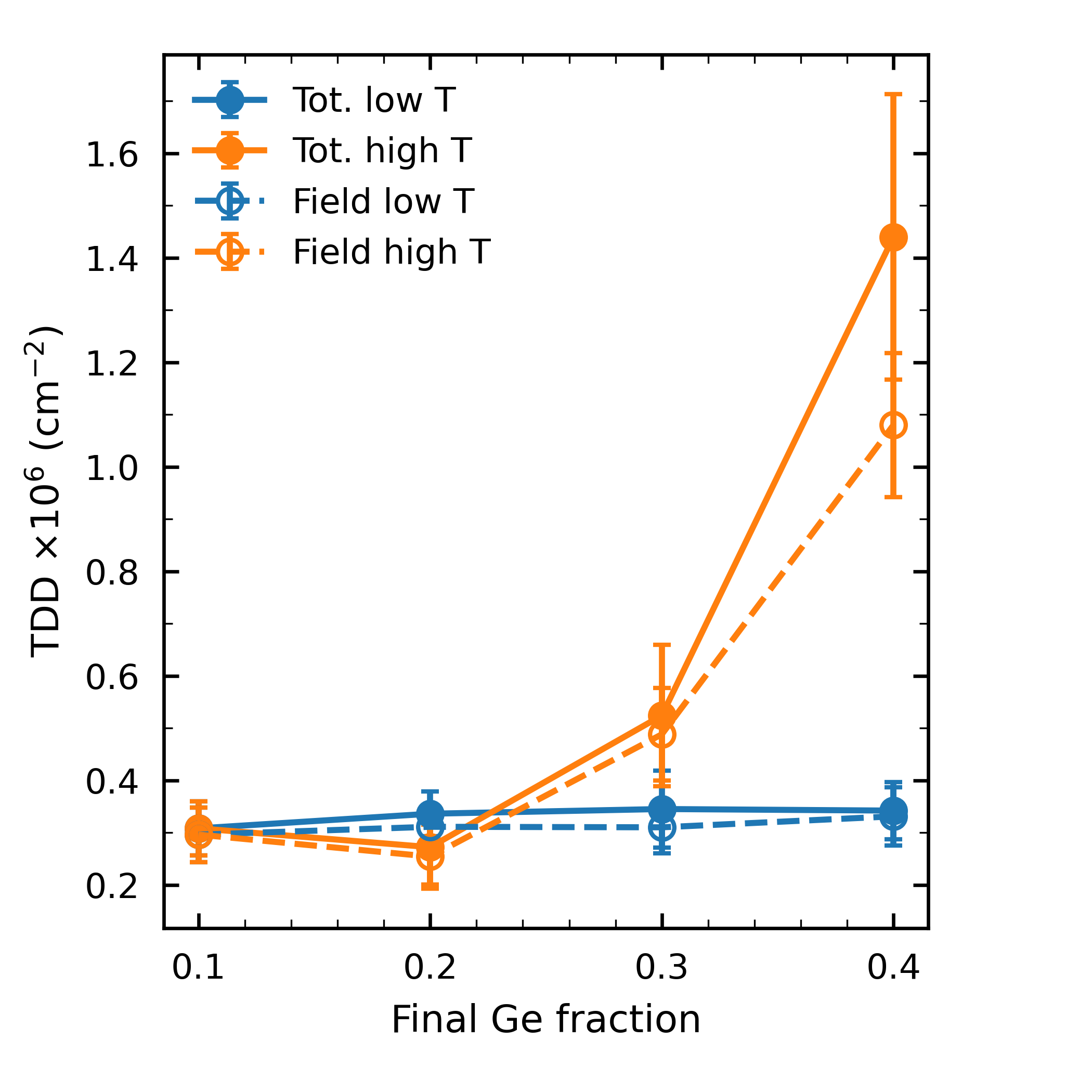}
    	\caption{\label{fig:TDDSliced} Total (filled circles) and field (empty circles) density of TDs in Si$_{1-x}$Ge$_{x}$ LGBs with increasing Ge content. The low T data-points refer to samples A1, B1, C1, and D, the high T data-points to A5, B5, and C5 described in  Tab.\,\ref{tab:samples}. }
    \end{figure}
The etch-pit counting and VT-HRXRD measurements consistently support the idea that, after the initial phase of the deposition where dislocation are nucleated in both the low-T (A1) and high-T (A5) samples two different relaxation mechanisms take place: in A1 the relaxation proceeds mainly by gliding of existing TD, while in A5 new dislocations are nucleated. 
To confirm this picture the TDD and degree of relaxation of samples with lower Ge content (B, C and D in Tab.\,\ref{tab:samples}) have also been investigated.

Figure\,\ref{fig:TDDSliced} shows the etch-pit counting for the low-T (A1, B1, C1 and D) and high-T (A5, B5, C5, D) Si$_{1-x}$Ge$_{x}$ LGBs series with $x_f= 0.4, 0.3, 0.2, 0.1$. It can be noticed that, for the high-T series the TDD increases for $x_f \succsim 0.3$ with an increased contribution from piled-up dislocations. In the low-T case the TDD is, instead, constant in the $x_f=0.1-0.4$ range in agreement with what expected for glide-driven relaxation \cite{abrahams_dislocation_1969}.
\captionsetup[subfigure]{
	position=bottom,
	skip=-20pt,
	singlelinecheck=false,
	justification=raggedright
}
\begin{figure}
	\centering
    \subcaptionbox{}
{\includegraphics[width=0.45\textwidth]{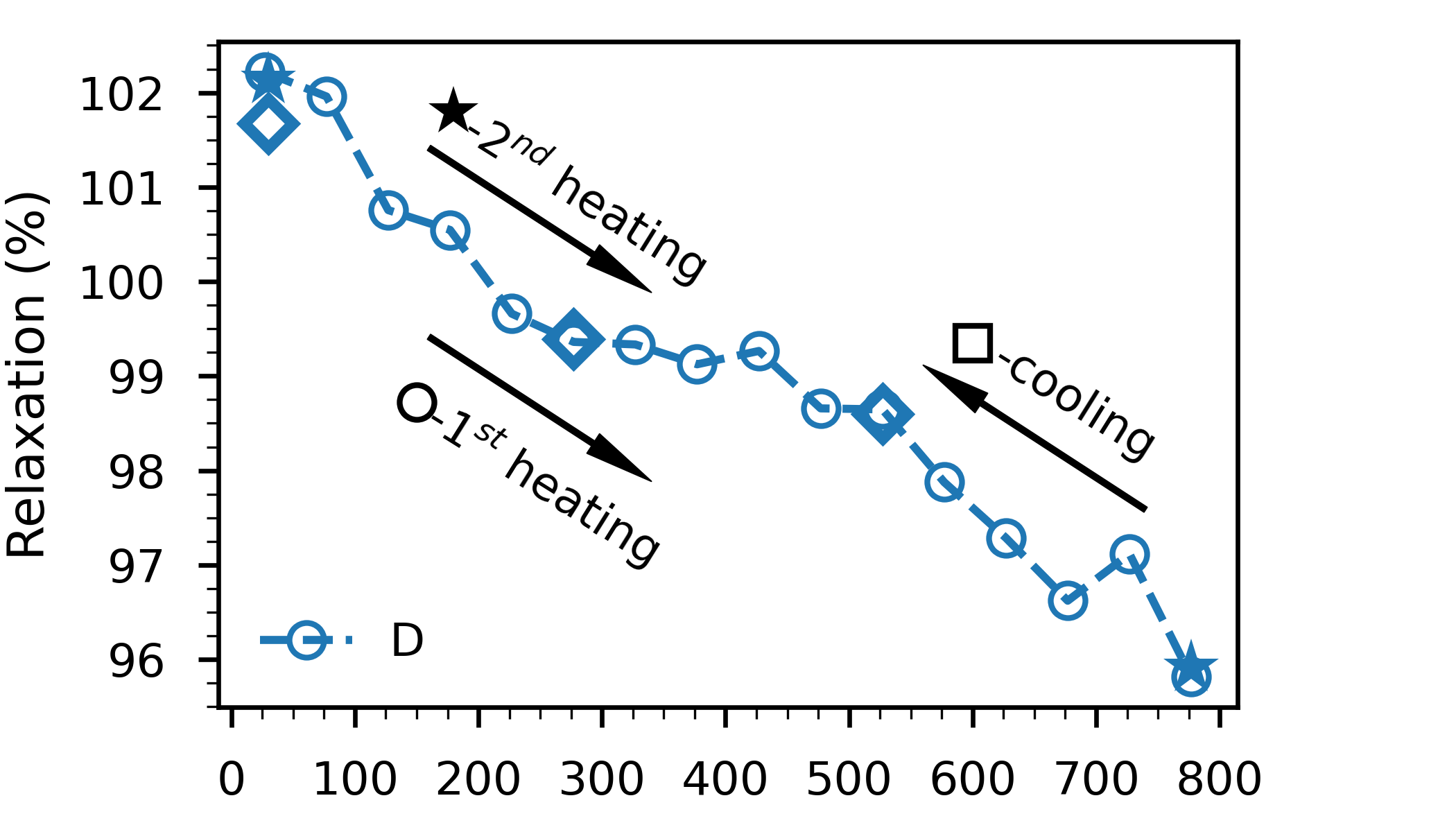}}
	\subcaptionbox{}
	{\includegraphics[width=0.45\textwidth]{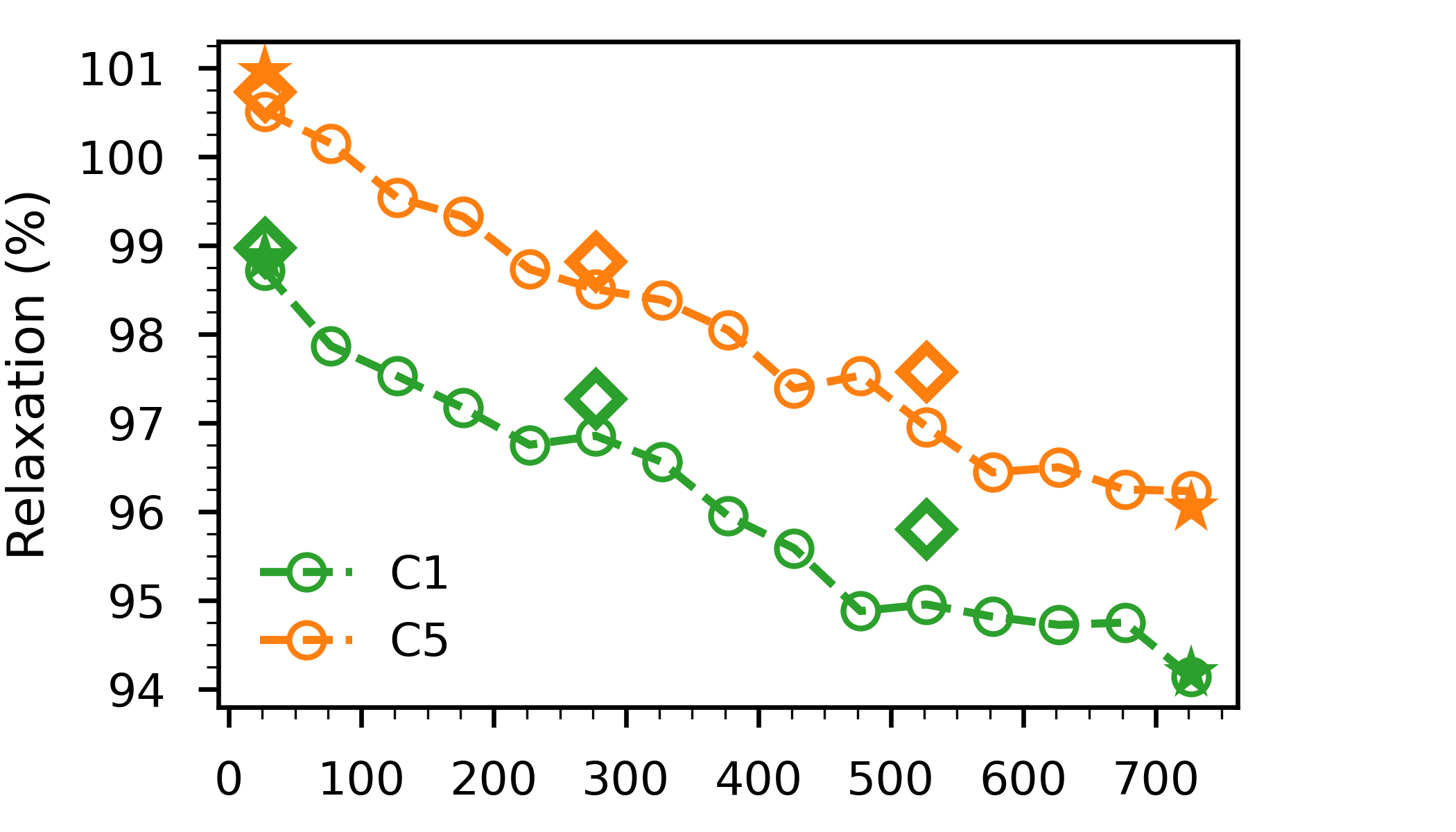}}
	\subcaptionbox{}
	{\includegraphics[width=0.45\textwidth]{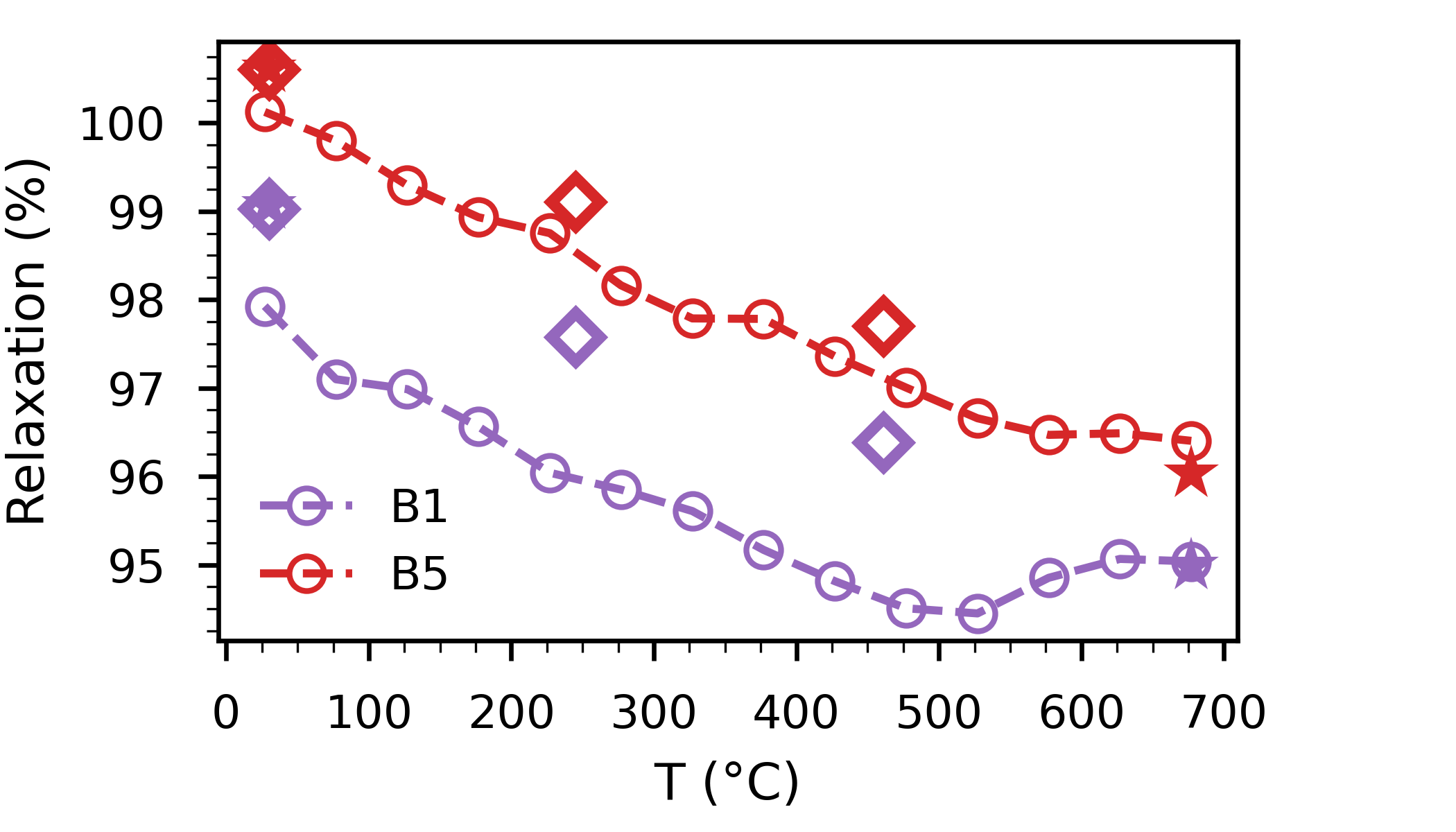}}
	\caption{Degree of relaxation at different annealing temperatures for Si$_{1-x}$Ge$_{x}$ LGBs with increasing Ge content. Relaxation of (a) sample D measured in the temperature range T=27-777$^\circ$C (b) samples C1 and C2 for T=27-727$^\circ$C and (c) B1 and B5 for T=27-677$^\circ$C.}	
	\label{fig:RelaxSliced}	
\end{figure}

VT-HRXRD data reported in Fig.\,\ref{fig:RelaxSliced} confirm that for samples D ($x=0.1$), C5 and C1 ($x=0.2$) no sizable change in the plastic relaxation of the misfit strain takes place, upon annealing. On the other hand, a comparison between the VT-HRXRD data of samples B1 ($x=0.3$ and T$_f$=580$^\circ$C) and B5 ($x=0.3$ and T$_f$=627$^\circ$C) shows that additional plastic relaxation, although less pronounced than in the case of sample A1, takes place in B1 for T$\succsim$ 500$^\circ$C.

The relatively narrow temperature range, where the transition from glide- to nucleation-driven relaxation occurs, suggests a possible correlation between our observations and the brittle to ductile transition (BDT) extensively studied in Si \cite{j_samuels_brittleductile_1989, p_b_hirsch_brittle-ductile_1989} and Ge \cite{serbena_brittle--ductile_1994} by means of fracture tests performed on indented samples. These studies highlighted a remarkable correlation between the BDT temperature T$_c$, above which several dislocations are nucleated at the indented crack tip, and the activation energy for dislocation glide $U$. For a wide range of materials \cite{hirsch_comment_1996} $U/k_BT_c\sim25$. Taking the approximate values of 1.5 and 2 eV for $U$ in Ge and Si \cite{hirsch_comment_1996}, the T$_c$ of the two materials can be linearly interpolated to obtain a rough estimation of the BDT temperature of Si$_{1-x}$Ge$_x$ alloys as shown by the gray region in  Fig.\,\ref{fig:TProfile}. 

It can be noticed that, during the growth of samples A5, the deposition temperature is always above T$_c$ favoring the nucleation of new dislocations. On the contrary, in the last phases of sample A1  the deposition temperature falls below T$_c$ inhibiting the nucleation of more dislocations and leaving glide as the only active relaxation mechanism. Although the comparison between fracture tests and epitaxial growth must be taken with care, our data might open up new perspective for the study of the role played by the BDT in hetero-epitaxy.
     
\section{Conclusions}
The relaxation of Si$_{0.6}$Ge$_{0.4}$ LGBs  has been investigated in a temperature range hardly accessible by thermal CVD, by growing, thanks to LEPECVD,  a set of samples with identical Ge composition but different temperature profiles. Defect-etching and VT-HRXRD  indicate that the mechanism driving misfit relaxation switches from the gliding of existing TDs for samples deposited at a final  T$\precsim$ 520$^\circ$C, to the nucleation of new dislocations for T $\succsim$ 530$^\circ$C. This is confirmed by a set of samples with lower Ge content featuring a composition-independent TDD in the $x=0.1-0.4$ range, a clear signature of glide-driven relaxation. Interestingly, a sharp increase of TDD was previously reported in Si$_{0.04}$Ge$_{0.96}$ LGBs deposited by thermal CVD \cite{isaacson_deviations_2006} at T$\geq 550$°C suggesting such a behavior to be independent to the specific growth conditions. A tentative explanation of these results can be given in terms of the brittle-to-ductile transition. This opens the way for further studies on such ubiquitous phenomenon, exploiting LGBs as an environment with controlled and tuneable strain-levels.

\begin{acknowledgments}
Part of this work has been carried out within the International Joint Lab “Intelligent Electro-Optical Sensing,” established between IHP - Leibniz Institute for High Performance Microelectronics and the University of Rome Tre.
A. C. D. acknowledges the support by the European Union’s Horizon Europe LASTSTEP Project under the grant agreement ID: 101070208.
D. I. acknowledges the support by the European Union’s ERC ELECTROPHOT Project, under the Grant Agreement ID: 101097569.
\end{acknowledgments}

\section*{Data Availability Statement}
The data that support the findings of this study are openly available in ZENODO, at \url{https://zenodo.org/records/17750768}, reference number 17750768.
\bibliography{MyBilbiography}

\end{document}